\documentclass[aps,pra,reprint,superscriptaddress,amsmath,amsfonts,amssymb]{revtex4-1}

\usepackage{graphicx}
\usepackage{hyperref}
\usepackage{amsthm}
\usepackage{qcircuit}
\usepackage{physics}
\usepackage{xstring}

\newcommand{\uba}{Departamento de F\'\i sica, FCEyN, UBA, Pabell\'on 1,
  Ciudad Universitaria, 1428 Buenos Aires, Argentina}
\newcommand{\ifiba}{Instituto de F\'\i sica de Buenos Aires, UBA CONICET,
  Pabell\'on 1, Ciudad Universitaria, 1428 Buenos Aires, Argentina}

\newcommand{\beq}{\begin{equation}}
\newcommand{\eeq}{\end{equation}}
\newcommand{\bse}{\begin{subequations}}
\newcommand{\ese}{\end{subequations}}
\newcommand{\bea}{\begin{eqnarray}}
\newcommand{\eea}{\end{eqnarray}}

\newcommand{\Wext}{W_{\text{ext}}}
\newcommand{\Wer}{W_{\text{er}}}

\newcommand{\Qer}{Q_{\text{er}}}

\newcommand{\Emeas}{\Delta E}
\newcommand{\Emeasj}[1]{\Delta E_{#1}}

\usepackage{xcolor}

\begin{document}

\title{Enhancing the efficiency of quantum measurement-based engines with entangling measurements}

\author{Franco Mayo}
 \email{fmayo@df.uba.ar}
 \affiliation{\uba} \affiliation{\ifiba}
\author{Augusto J. Roncaglia}
  \email{augusto@df.uba.ar}
  \affiliation{\uba} \affiliation{\ifiba}


\begin{abstract}
   We study the impact of entangling measurements on the efficiency of quantum measurement-based engines. We first show that for engines comprising many subsystems their efficiency can be enhanced by performing  entangling measurements, as opposed to local measurements over each subsystem. When the collective measurement produces the same local state for the subsystems as individual local measurements, the improvement in the efficiency is proportional to the amount of correlations. Finally, we show that for two level systems these type of engine can operate at perfect efficiency while yielding a finite amount of work, in the limit large the number of subsystems. 
    
\end{abstract}

\maketitle  

\section{Introduction}
\label{sec:intro}
As thermal machines drove the development of classical thermodynamics, quantum thermodynamics has gained interesting insights from the study of quantum heat engines and refrigerators~\cite{binder2019thermodynamics, vinjanampathy2016quantum}. For instance, we can mention the development of shortcuts to adiabaticity~\cite{torrontegui2013shortcuts} and shortcuts to equilibrium~\cite{dann2019shortcut, pancotti2020speed}, that allow thermal machines to increase their power without loosing efficiency, or the role that quantum features play in thermodynamics. The study of quantum thermal machines includes autonomous thermal machines~\cite{linden2010small,brunner2012virtual}, heat engines operating with quantum coherence~\cite{dann2020quantum, camati2019coherence, hammam2021optimizing} and quantum correlations~\cite{hewgill2018quantum}. Also, experimental realizations of heat engines on the quantum scale have been made~\cite{rossnagel2016single,von2019spin}.

In recent years, the idea of quantum engines in which the fuel is not heat from a thermal reservoir but quantum measurements performed on it arose~\cite{jordan2020quantum}. These engines take advantage of the fact that performing some measurement on a system can change its energy, provided that the measurement does not commute with the system's hamiltonian. In this way, the heat provided by a thermal reservoir can be replaced by the energy supplied by a measurement. Several quantum measurement engines (QMEs) have been introduced during the last years; these include single qubit engines~\cite{elouard2017extracting}, a quantum Otto engine in which the hot bath is replaced by a non-selective quantum measurement~\cite{yi2017single}, a quantum Carnot engine~\cite{fadler2023efficiency}, a climbing particle assisted by a quantum measurement~\cite{elouard2018efficient}, a measurement engine fueled by entanglement~\cite{bresque2021two} and a refrigerator~\cite{buffoni2019quantum}. In addition, there were also different proposals that use incompatible measurements to fuel quantum measurement engines~\cite{opatrny2021work, manikandan2021efficiently} and other engines that work without using feedback~\cite{das2019measurement,ding2018measurement}, as well as works considering multilevel systems~\cite{de2021efficiency,anka2021measurement} or interacting qubits~\cite{bhandari2023measurement}. 
The thermodynamics of quantum measurements have been also studied for instance by considering the cost to perform quantum measurements~\cite{guryanova2020ideal,sagawa2009minimal,jacobs2009second,jacobs2012quantum,abdelkhalek2016fundamental,deffner2016quantum,linpeng2022energetic, perna2023fundamental}, the role of quantum coherence~\cite{lin2021suppressing} and the classicality of the heat produced during measurements~\cite{mohammady2021classicality}. 
These machines have some very interesting properties compared to traditional thermal machines. 
For instance, its efficiency is not bounded by Carnot's efficiency, as there is only one temperature to consider, and the maximum efficiency can be achieved at maximum power~\cite{elouard2017extracting}. 
This is in contradiction to what happens to thermal machines, where there is a trade-off between power and efficiency. 

In this work we aim to look at an interesting feature of quantum mechanics that might lead to interesting results in the realm of quantum thermodynamics, collective entangling measurements.  We show that if one considers QMEs composed of many subsystems, it is possible to obtain a higher efficiency when collective entangling measurements as opposed to local ones over each individual subsystem. We also find that perfect efficiency can be achieved in the limit of a large number of subsystems.
\begin{figure}[tb]
    \centering
    \includegraphics[width=0.4\textwidth]{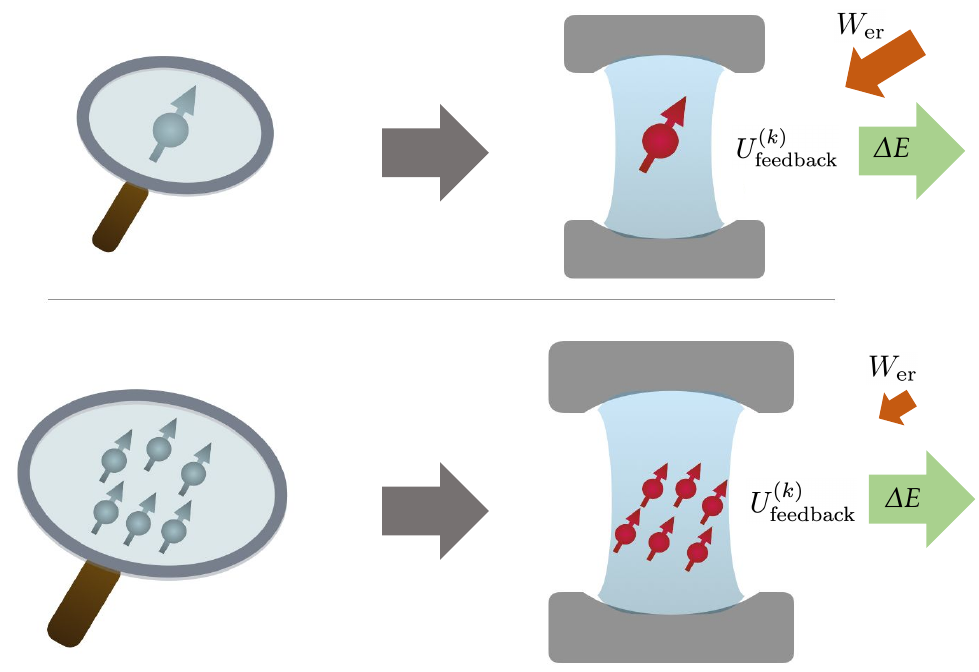}
    \caption{Schematic representation of the (top) individual or parallel and (bottom) collective process. The QME cycle starts with a measurement that increases the system's energy. Afterwards, work is extracted by performing a feedback operation $U_{\rm feedback}$ leaving the system in its initial state. The cycle ends after erasing the information about the measurement result. In the parallel process both measurement and feedback (work extraction) are performed independently on each system, while in the collective one both operations are performed globally on all systems. As indicated by the size of the arrows, the collective process can lead to smaller erasure cost.}
    \label{fig:collective_schema}
\end{figure}

The paper is organized as follows. In Sec.~\ref{sec:efficiency_of_QME} we introduce quantum measurement engines and discussed how its efficiency is defined. Afterwards in Sec.~\ref{sec:collective_measurements} we show how the use of global entangling measurements enhances the efficiency of QMEs. Finally, in Sec.~\ref{sec:conclusions} we present some discussions and the conclusions.

\section{Efficiency of quantum measurement engines}
\label{sec:efficiency_of_QME}

Let us first recall how a typical measurement engine works. The working body consists of a system $\mathcal{S}$ with hamiltonian $H_\mathcal{S}$ that starts in a given pure state, and there is an environment at equilibrium with temperature $T$. Thus, the QME can be described in terms of the following steps:

\emph{(a) Projective measurement}: Initially, the system is in a pure state $\ket{\psi_{\rm i}}$, with energy $E_{\rm i} = \ev{H_\mathcal{S}}{\psi_{\rm i}}$, and is measured on some basis $\{\ket{k}\}$ that does not commute with $H_\mathcal{S}$. After the measurement, the system is left in the state $\ket{k}$ with probability $p_k= \abs{\braket{k}{\psi_{\rm i}}}^2$, and its averaged energy (considering all possible measurement outcomes) is ${E_{\rm f} = \sum_k p_k \ev{H_\mathcal{S}}{k}}$. If the measurement basis is chosen properly, then $E_{\rm f}>E_{\rm i}$ and the measurement has increased the systems energy by $\Emeas = E_{\rm f} - E_{\rm i}$.

\emph{(b) Feedback \& work extraction}: If the outcome of the measurement was $\ket{k}$ the system is restored to the initial state $\ket{\psi_{\rm i}}$ with some unitary operation $U_{\text{feedback}}^{(k)}$.  During this process an amount of work $W_k = \ev{H_\mathcal{S}}{k} - \ev{H_\mathcal{S}}{\psi_{\rm i}}$ can be extracted. On average, the amount of extracted work is $\Wext = \Emeas$.

\emph{(c) Erasure}: The cycle ends once the classical memory that stores the result of the measurements is erased. According to Landauer's principle~\cite{landauer1961irreversibility} this can be done quasi-statically using some work $\Wer$ that is equal to the amount of heat ${\Qer = k_B T S_{\rm f}}$ dissipated to the environment, and $S_{\rm f}$ is the Shannon entropy of the memory. This entropy is also equal to the von Neumann entropy of the system after the measurement ${S_{\rm f}\equiv H\{p_i\}=-\sum_k p_k\log p_k}$.

The aforementioned cycle is designed in such a way that, on average, the system absorbs an amount of energy $\Emeas$ from the measurement, while the output work is ${W=\Wext-\Wer=\Emeas-\Wer}$. Thus, in general it works like an engine, provided the temperature of  the reservoir  is low enough so that $\Emeas>\Wer$.

Given that the measurement generally leads to an increase of the system entropy, this form of energy is commonly referred to as `quantum heat'~\cite{elouard2017role}. However, in other frameworks, this energy is referred to as work~\cite{sagawa2009minimal,jacobs2009second} since it can leave the system in a non-equilibrium state. We agree with this second interpretation, associated to the external work required to couple the system with a measurement apparatus.
In this way, the rise in entropy is just a direct consequence of the correlations established during the interaction. Hence, we can interpret these engines as transducers of energy from measurements.
Thus, the efficiency of this cycle is 
\beq
\eta=\frac{W}{\Emeas}.
\eeq
Notably the efficiency can attain values close to one~\cite{elouard2017extracting, elouard2018efficient}, for instance, when the outcomes of the measurement have low dispersion. However, in this situation the amount of extracted work is vanishing small, so there is a trade-off between efficiency and extracted work. 

 In order to analyze the performance of these engines in more general scenarios, we will introduce an alternative way of writing the efficiency. Since the net work is $W=E_{\rm f}-E_{\rm i}- k_BT S_{\rm f}$, it is easy to notice that efficiency can also be expressed as: $\eta={\Delta F}/{\Delta E}$, where $F=\expval{E}- k_BT S(\rho)$ is the free energy, and differences are taken between averaged states after and before measurement, that is: $\rho_{\rm f}=\sum_k p_k \ketbra{k}{k}$ and $\rho_{\rm i}\equiv\ketbra{\psi_{\rm i}}{\psi_{\rm i}}$, respectively, and $\Delta E = \tr[H_\mathcal{S} \rho_{\rm f}]-\tr[H_\mathcal{S} \rho_{\rm i}]$ is the difference in the mean energy between the same states.  As we know, $\Delta F$ is also the maximum (minimum) work that can be extracted (yielded) in the transformation $\rho_{\rm f}\rightarrow \ketbra{\psi_{\rm i}}{\psi_{\rm i}}$ in contact with a heat bath at temperature $T$. Therefore, one could disregard the specific feedback or transformation implemented to extract work and solely focus on the maximum work that can be extracted during the transformation from the post-measurement state to the initial one. This shows that the energetics involved in a QME is equivalent to a two-step cycle where: $(i)$ a decoherence process (in the measurement basis) over the initial state, and  $(ii)$ a transformation in contact with a heat bath  that restores the system to its initial state, during which some work is extracted. 

As we shall see, this picture is useful as it enables us to explore more general QME, and the straightforward evaluation of their efficiency in situations involving collective measurements.

 \section{Collective measurements}
\label{sec:collective_measurements}
So far we have discussed engines in which there is a single quantum system. We now consider the case where there is an engine consisting of $N$ subsystems where the $N-$QME Hamiltonian is $H=\sum_{\mu=1}^N h_\mu$, and focus on the difference between performing local measurements and  collective measurements, as it is sketched out in Fig.~\ref{fig:collective_schema}. 
The aim of this section is to show that it is possible to improve the performance of these engines by implementing collective measurements. In this way one can enlarge the working range of the engine, extracting more work at larger efficiencies. 

Let us first consider a QME running in parallel, that is each measurement is done locally. In this case, on average the energy provided by the measurement is just 
${\Emeas}=\sum_{\mu} \Emeasj{\mu}$, where 
$\Emeasj{\mu}=\tr[h_\mu(\rho_{{\rm f},\mu}- \rho_{\rm i,\mu})]$ 
is the energy provided by each single engine. Then, if $W_\mu$ is the net extracted work of each QME, one can define the efficiency of the parallel process as $\eta_{\|} =\sum_\mu W_{\mu}/\sum_{\mu} \Emeasj{\mu} \equiv 1- k_BT\sum_\mu S_\mu/\sum_\mu \Emeasj{\mu} $, where $S_\mu$ is the entropy generated by the $\mu-$th engine whose post measurement state is $\rho_{{\rm f},\mu}$. 

Now, one can compare this efficiency with the efficiency of an engine subjected to a collective measurement that outputs the same mean energy. For instance, this can be done by imposing that the post-measurement $N$-partite state $\rho_{\rm f}$ is such that is locally equivalent to the one associated with the parallel process:
\begin{equation}
    \Tr_{- \mu}[\rho_{\rm f}] = \rho_{\rm f, \mu}
\end{equation}
this ensures that on average the energy provided by the measurement is the same in both processes. 
 
Then, the efficiency of the collective process is defined as:
    $\eta_\sharp = 1 - k_B{T S_\sharp}/{\sum_\mu \Emeasj{\mu}}$,
where $S_\sharp$ is the entropy generated by the collective measurement. Now we can express this entropy in terms of  the multipartite mutual information~\cite{watanabe1960information, kumar2017multiparty} as 
$S_\sharp = \sum_{\mu}^N S_\mu -\mathcal{I}$,

where $S_\mu$ is the entropy of each subsystem and $\mathcal{I}$ is the multipartite mutual information, that accounts for all the correlations created during the measurement. 
This allows us to write the efficiency of the collective process as:
\begin{equation}
    \eta_\sharp = \eta_{\|} + \frac{k_BT\ \mathcal{I}}{\sum_\mu \Emeasj{\mu} }.
\end{equation}
Thus, we can see that performing a collective measurement that generates correlations between the subsystems enhances the efficiency of the QMEs. 
It is worth noting that the last expression does not indicate an increase in the efficiency with temperature. Considering that the first term is also temperature-dependent, the efficiency could potentially be negative at high temperatures. Hence,  it is not straightforward to determine the significance of this improvement. Below, we will show that in the  symmetric QME scenario, the efficiency could be close to one while still extracting a finite amount of net work per subsystem.

\subsection{Symmetric QME}

We will consider a \emph{symmetric QME},  that is an engine composed by identical subsystems, i.e. they start in the same initial state, are subjected to the same measurement, and have the same hamiltonian. In this case, it is easy to see that ${\sum_\mu W_\mu=N W_1}$, ${\Emeas=N \Delta E_1}$, $\eta_{\|}=\eta_1$, and therefore ${\eta_\sharp=\eta_{\|} +k_BT \mathcal I /(N \Delta E_1)}$ (where the subindex 1 indicates that the quantity is associated to a single QME). 

Let us start by considering a simple example, an engine composed by two two-level systems. 
When this engine runs in parallel, each system is measured on a given basis that we call $\{\ket 0, \ket 1\}$, and starts in a general pure state that can be written in this basis as ${\ket{\psi_{\rm i}} = \sqrt q \ket{0} + \sqrt p \ket{1}}$, where we choose positive real numbers without loss of generality. 
Therefore, if the Hamiltonian for a single system is $h_1$,
$E_{\rm i}=q\bra{0}h_1 \ket{0}+p\bra{1}h_1\ket{1}+2\sqrt{qp}\ \mathrm{Re}\{\langle 0 \vert h_1 \vert 1 \rangle\}$, and it has an efficiency $\eta_1=1 - k_BT S_1 /{\Emeasj{1}}$, where 
$S_1 =-q\log q-p\log p$ and
$\Emeas_1 = -2\sqrt{qp}\ \mathrm{Re}\{\langle 0 \vert h_1 \vert 1 \rangle\}$.

Now we will show that by using a collective measurement that generates correlations between both systems, one can fuel the engine with the same amount of energy as in the parallel process, but with a higher efficiency. In this case the initial state is $\ket*{\psi^{(2)}_{\rm i}} \equiv \ket{\psi_{\rm i}}\otimes\ket{\psi_{\rm i}}$, that  can be written as:
\begin{equation}
    \ket{\psi^{(2)}_{\rm i}} = q\ket{00} + p\ket{11} + \sqrt{qp}\  ({\ket{01} + \ket{10}})
\end{equation}
For the collective measurement basis we make the following choice: $\{\ket{00}, \ket{11},\frac{\ket{01} + \ket{10}}{\sqrt{2}}, \frac{\ket{01} - \ket{10}}{\sqrt{2}}\}$. It is easy to see that for this choice, the post-measurement state is 
$\rho^{(2)}_{{\rm f},\sharp} = p^2 \ketbra{00}{00} + q^2 \ketbra{11}{11} + pq(\ket{01}+\ket{10})(\bra{01}+\bra{10})$, which is locally equivalent to that of the parallel process: $\rho_{{\rm f},1}=\rho_{{\rm f},2} = q\ketbra{0}{0}+p\ketbra{1}{1}$.
Therefore, both the local and the collective measurements will increase the systems' energy by the same amount: ${\Emeas^{(2)}=2 \Emeas_1}$. However, it is interesting to notice that the probability of obtaining the last element of the measurement basis is zero, which in turn reduces the entropy of the post-measurement state. This 
entropy  is
${S^{(2)}_{\sharp} = 2 (-q\log q - p\log p) - 2p q\log 2= 2 S_1 - 2p q\log2}$, i.e, $S^{(2)}_{\sharp} \leq 2 S_1$, and the net extracted work is: 
 \beq
 W_\sharp^{(2)}=W_{\|}^{(2)}+ qp  \frac{ k_B T \log2}{\Delta E_1}.
 \eeq
Therefore, since this example is valid for an arbitrary local measurement, it tells us that for every parallel engine of this type, there exists a collective measurement that enhances its efficiency. Naturally, this also requires  a collective unitary operation as a feedback.

In the following, we will generalize this idea for an arbitrary number $N$ of systems. 
For $N$ two-level systems, the initial state is associated to the following tensor product:
\beq
    \ket{\psi_{\rm i}}^{\otimes N} = \sum_i \sqrt{q^{(N-i)}p^{i}}\ \mathbf{S}\left(\ket{0}^{\otimes {N-i}} \ket{1}^{\otimes i}\right),
    \label{eq:initial_state_N}
\eeq
where $\mathbf{S} = \sum_\pi P_\pi$
is a symmetrization operator and $P_\pi$ is a permutation operator running over all permutations $\pi$. 
In this way, we can see that the state is a linear combination of $N$ orthonormal vectors:
$\ket{i_\sharp}\equiv \mathbf{S}\left(\ket{0}^{\otimes {N-i}} \ket{1}^{\otimes i}\right)/\sqrt{{N \choose i}}$ labelled by the index $i$, which can be associated to the Hamming weight or the ``occupation number''. These vectors span a subspace of the Hilbert space of the $N$ systems $\mathcal H^{\otimes N}$ that is invariant under all permutations, and we call it $\mathcal H_+^{\otimes N}$.
Now, we can define a collective measurement basis that contains the $N$ elements $\{\ket{i_\sharp}\}$ 
expanding $\mathcal H_+^{\otimes N}$, and the remaining elements are composed by an arbitrary basis expanding the orthogonal projection of $\mathcal H_+^{\otimes N}$ to $\mathcal H^{\otimes N}$.
In this way, one can show that the post-measurement state is
\begin{equation}
    \rho_{{\rm f},\sharp} = \sum_i {N\choose i} \ q^{N-i}p^{i} \ \ketbra{i_\sharp}{i_\sharp}.
    \label{eq:binomial_state}
\end{equation}

Where the probability associated to the different outcomes are determined by a binomial distribution. Therefore, the change in entropy after the measurement is just the entropy of the binomial distribution, whose expression for a large $N$ is~\cite{knessl1998integral}:
\begin{equation}
    S_\sharp = \frac{1}{2}\,\log(2\pi e N qp) + \mathcal{O}\left(\frac{1}{N}\right).
    \label{eq:binomial_entropy}
\end{equation}
If we compare this situation with the local or parallel measurement scheme, in that case the entropy of the measurement is $S_\|=  N S_1$, while in the global measurement scenario it scales as $S_\sharp \sim \log N$. However, 

it is easy to see that the energy provided by the measurement is the same in both cases:
given that $\bra{i_\sharp}\sum_j h_{\mu}\ket{i_\sharp} = (N-i)\bra{0}h_1\ket{0}+i \bra{1}h_1\ket{1}$, and using Eq.~\eqref{eq:binomial_state}  it is easy to verify that $E_{\rm i} = N (q\bra{0}h_1 \ket{0}+p\bra{1}h_1\ket{1})$. Thus, as $\Emeas=N\Emeas_1$, the efficiency of the collective engine is therefore
\begin{equation*}
    \eta_\sharp = 
    1 - k_BT \frac{\log (2\pi e N p q) /2 + \mathcal{O}(\frac{1}{N})}{N \Emeas_1},
\end{equation*}
Therefore, we can see that for large $N$ it scales as:
\begin{equation}
    \eta_\sharp \sim 1 - \frac{k_BT}{\Emeas_1} \frac{\log N}{N},
    \label{eq:scaling_log}
\end{equation}
and one can approach $\eta_\sharp\xrightarrow[N \to \infty]{} 1$. 

\begin{figure}[tb]
    \centering
    \includegraphics[width=0.48\textwidth]{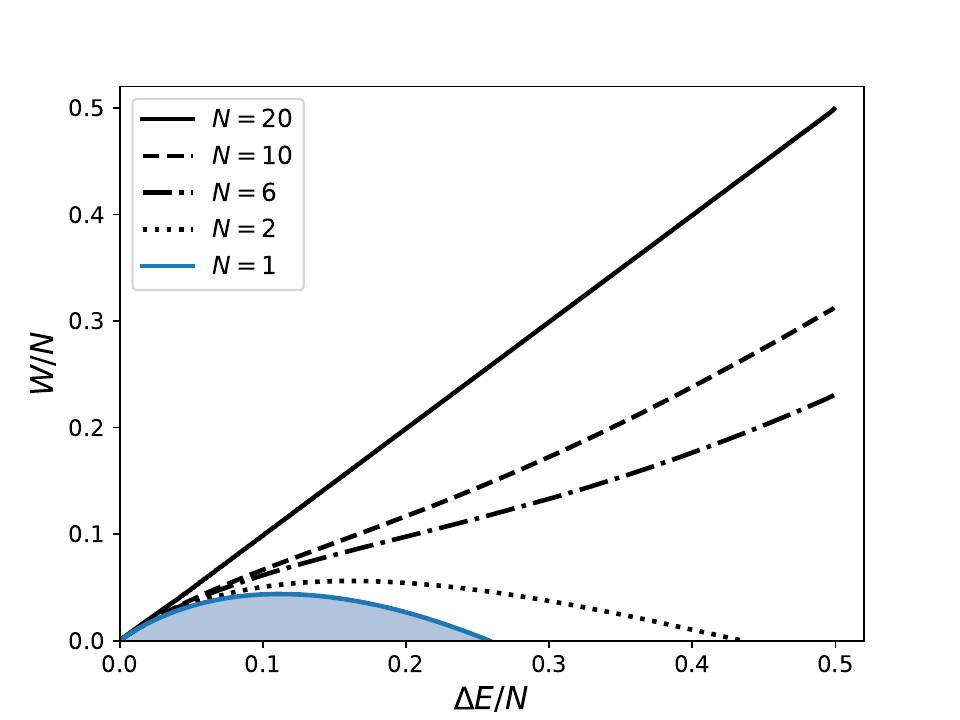}
    \caption{Extractable work as a function of the energy provided by the measurement $\Emeas/N =\Emeas_1$ for different number of two-level systems. The shaded blue area depicts the full range of single two-level QMEs, where each point corresponds to a specific initial state and measurement basis. The curves correspond to the collective strategies that are locally equivalent to the optimal parallel QMEs, represented by the blue curve. Results are shown for $N=1$ (blue), $N=2$ (dotted), $N=6$ (dot-dashed), $N=10$ (dashed) and $N=20$ (solid). We can observe that as $N$ grows so does the efficiency of these engines. In fact, for $N=20$ the slope of the curve is almost one, indicating almost perfect efficiency.}
    \label{fig:eff_vs_N}
\end{figure}

Quantum measurement engines have been show to approach efficiency one in~\cite{elouard2017extracting}, however, this was done at the expense of obtaining a vanishing amount of work. In this case, we find that for every local measurement, there is a collective one that allow us to extract the same amount of work (per system) with unit efficiency in the large $N$ limit. In Fig.~\ref{fig:eff_vs_N} we show that one can surpass the efficiency of every possible single two-level QME, and  even for a small number of systems one can attain efficiency one for every local measurement. 

It is interesting to notice that the above strategy works for any initial pure state and local measurement basis,
however it is not optimal.
For instance, one can consider $N$ two-level systems starting in the ground state $\ket*{\psi_{\rm i}^{(N)}} = \ket{g}^{\otimes N}$, and the following collective measurement basis: 
\begin{equation*}
    \mathcal B_{\sharp}=\left\{\frac{\ket{g}^{\otimes N} + \ket{e}^{\otimes N}}{\sqrt{2}}, \frac{\ket{g}^{\otimes N} - \ket{e}^{\otimes N}}{\sqrt{2}}, \mathcal{C}\right\},
\end{equation*}
consisting of two orthogonal GHZ states along with a basis  $\mathcal C$ that expands their orthogonal complement. It is easy to see that the equivalent parallel process is the one where each system is measured in the local basis $\{\frac{\ket{g}+ \ket{e}}{\sqrt{2}}, \frac{\ket{g}- \ket{e}}{\sqrt{2}}\}$ where the post-measurement state of each subsystem is proportional to the identity. It is easy to see that the same happens for the collective strategy.  
In this case, the measurement has only two possible outcomes regardless of the value of $N$, and both outcomes happens with the same probability. Therefore the entropy after the measurement is always $S_\sharp=\log 2$. Thus, the efficiency of this engine is: 
\begin{equation*}
    \eta_\sharp= 1 - \frac{k_B T }{\Emeas_1}\frac{\log 2}{N}.
\end{equation*}
In particular, this scaling is optimal, since the generated entropy does not increase with the number of systems.

Therefore, we have found a way of enhancing the efficiency of measurement engines by using collective measurements while maintaining the energy of the measurement constant. This enhancement happens for every initial state and choice of local measurement basis, but is not necessarily optimal. We also define a process in which the optimal collective advantage is achieved. 
Measuring multiple two level systems in a collective way is not realistic in practice; however, we find that there are considerable enhancements in the efficiency even for the case of two two-level systems, a scenario that has been realized experimentally several times, for example with measurements in the Bell basis~\cite{welte2021nondestructive}. Also, in order to obtain the efficiency enhancement one must perform the feedback globally, i.e. using a global unitary over all the systems.  Finally, it should be noted that for collective measurements one can obtain perfect efficiency in the limit $N\rightarrow\infty$, i.e. the process becomes reversible.  
In this case the energy supplied by the measurements is more work-like, and it can be described as a transducer of the measurement energy.

\section{Conclusions}
\label{sec:conclusions}

We have studied the behaviour of quantum measurement based engines under collective measurements.
We have shown that this type of measurements can generate correlations between the subsystems, enhancing the efficiency the cycle as they are able to reduce the entropy production. In this way, it is possible to obtain an efficiency as high as desired by adding subsystems, while extracting a finite amount of work. It has been shown in previous works that the efficiency of quantum measurement engines attain values close to one, in the limit of vanishing work extraction. Practical implementations of global measurements is challenging, however we have shown that a significant enhancement can still be achieved with a  small amount of systems. Our findings also shed light on the nature of the energy provided by the measurement.

\begin{acknowledgments}

This work was partially supported by CONICET, UBACyT (20020130100406BA) and ANPCyT (PICT-2019-04349 and PICT-2021-01288). 

\end{acknowledgments}

\end{document}